\documentclass[12pt]{article}
\usepackage{newtxtext,newtxmath}
\usepackage{graphicx}
\usepackage[letterpaper,margin=1in]{geometry}
\renewenvironment{abstract}
	{\quotation}
	{\endquotation}
\date{}
\usepackage{physics}
\usepackage{mathtools}

\usepackage{caption}
\usepackage{setspace}
\usepackage{tikz}
\usetikzlibrary{patterns}
\usetikzlibrary{shadows.blur}
\usetikzlibrary{shapes}
\usepackage{tabu}
\usepackage{booktabs}
\usepackage{booktabs,dcolumn,caption}

\makeatletter
\renewcommand{\fnum@figure}{\textbf{Figure \thefigure}}
\renewcommand{\fnum@table}{\textbf{Table \thetable}}
\makeatother
\usepackage{scicite}
\usepackage[T1]{fontenc}
\usepackage{url}
\def\scititle{
	Direct Observation of Massless Excitons and Linear Exciton Dispersion
}
\title{\bfseries \boldmath \scititle}
\newcolumntype{d}[1]{D{.}{.}{#1}} % "decimal" column type
\author{%
  Luna Y. Liu\textsuperscript{1},%
  Steffi Y. Woo \textsuperscript{1, 3‡},%
  Jinyuan Wu\textsuperscript{3},%
  Bowen Hou\textsuperscript{3},%
  Cong Su\textsuperscript{1, 3*},%
  Diana Y. Qiu\textsuperscript{3, 4†}%
  \and
  \small\textsuperscript{1}Department of Applied Physics, Yale University, New Haven, CT, 06511, USA.%
  \and
  \small\textsuperscript{2}Center for Nanophase Materials Sciences, Oak Ridge National Laboratory, Oak Ridge, TN, 37831, USA.%
  \and
  \small\textsuperscript{3}Department of Material Science, Yale University, New Haven, CT, 06511, USA.%
  \and
  \small\textsuperscript{4}Energy Science Institute, Yale University, West Haven, CT 06516, USA.%
  \and
  \small\textsuperscript{‡}Corresponding author. Email: woosy@ornl.gov.%
  \and
  \small\textsuperscript{*}Corresponding author. Email: cong.su@yale.edu.%
  \and
  \small\textsuperscript{†}Corresponding author. Email: diana.qiu@yale.edu.%
}

\begin{document} 
\maketitle
% Abstract, in bold
% There are strict length limits, and not all formats have abstracts.
% Consult the journal instructions to authors for details.
% Do not cite any references in the abstract.
\begin{abstract} \bfseries \boldmath
Excitons ---  elementary excitations formed by bound electron-hole pairs --- govern the optical properties and excited-state dynamics of materials. In two-dimensions (2D), excitons are theoretically predicted to have a linear energy-momentum relation with a non-analytic discontinuity in the long wavelength limit, mimicking the dispersion of a photon. This results in an exciton that behaves like a massless particle, despite the fact that it is a composite boson composed of massive constituents. However, experimental observation of massless excitons has remained elusive. In this work, we unambiguously experimentally observe the predicted linear exciton dispersion in freestanding monolayer hexagonal boron nitride (hBN) using momentum-resolved electron energy-loss spectroscopy. The experimental result is in excellent agreement with our theoretical prediction based on \textit{ab initio} many-body perturbation theory. Additionally, we identify the lowest dipole-allowed transition in monolayer hBN to be at 6.6 eV, illuminating a long-standing debate about the band gap of monolayer hBN. These findings provide critical insights into 2D excitonic physics and open new avenues for exciton-mediated superconductivity, Bose-Einstein condensation, and high-efficiency optoelectronic applications.
\end{abstract}

\newpage
% The first paragraph of any Science paper does NOT have a heading
% Nor is it indented
\noindent
The electronic band structure of a crystalline material defines the relationship between the allowed electron energies and momenta, fundamentally determining the electronic, optical, and magnetic properties of the material. The curvature of these bands can be interpreted as an effective electron mass, describing how an electron in a crystal responds to external forces. In rare instances, electrons can exhibit linear dispersion, and behave as if they are massless. For example, graphene and certain topological materials host Dirac fermions --- particles that follow the Dirac equation originally formulated for relativistic electrons \cite{novoselov2005two, hasan2021weyl, gomesdesigner, castro2009electronic,qi2011topological}. These Dirac fermions disperse linearly and maintain a constant velocity regardless of energy, similar to photons, allowing them to form wave packets with matching group and phase velocities. This unique behavior results in exceptional electronic properties, such as high electron mobility and conductivity, as well as resistance to specific scattering processes \cite{novoselov2005two, Liang_2014, geim2007rise}.

Similarly, excitons --- the elementary excitation of correlated electron-hole pairs --– possess their own bandstructure, which governs light-matter interactions and coupling to other electro-magnetic fields. Unlike electrons, excitons are composite bosons, which allows for the possibility of macroscopic quantum order in collective states like Bose-Einstein condensates and superfluids \cite{eisenstein2004bose, liu2022crossover}. Due to their composite nature, excitons usually exhibit a parabolic dispersion inherited from their constituent electron and hole. If excitons could exhibit a linear dispersion, it could lead to higher critical temperatures, stabilization of coherent phases, and the possibility of ultrafast ballistic energy transport \cite{tulyagankhodjaev2023}. Such a linear dispersion is in fact predicted to appear in 2D materials, where the reduced dimensionality of the long-range electron-hole exchange interaction gives rise to a linear dispersion with a non-analytic discontinuity in the exciton band structure \cite{louie2021discovering,qiu2015nonanalyticity,qiu2021signatures,yu2014dirac,wu2015}. This phenomenon is analogous to pions in particle physics, where composite particles formed by massive quarks and antiquarks appear nearly massless under certain conditions \cite{Franz1992}. The massless excitonic mode offers a potential condensed matter analog to such particle physics systems.

Despite the theoretical prediction of the linear exciton dispersion in 2D, direct measurement from experiment is challenging. Optical spectroscopy is limited by the small momentum of photons at the energy of typical material band gaps. Angle-resolved photoemission spectroscopy (ARPES) provides direct access to the electronic band structure, and recent developments in time-resolved ARPES (tr-ARPES) allow for the visualization of the spectral function of electrons photoemitted from excitons \cite{Andrea2024, man2021experimental, Alessandra2022}. However, while tr-ARPES provides access to the distribution of electron and hole momenta within the exciton, it cannot directly reveal the dispersion with the exciton center-of-mass momentum, $\mathbf{Q}$ \cite{rustagi2018photoemission,man2021experimental,karni2022structure}.  Techniques like momentum-resolved electron energy-loss spectroscopy (\textbf{Q}-EELS) and resonant inelastic X-ray scattering can probe both electronic and excitonic excitations \cite{Schuster2018, nicolaou2025}. There have been efforts to measure the excitonic band structures of monolayer transition metal dichalcogenides (TMDs) by \textbf{Q}-EELS \cite{Hong2020,Hong2021}. However, due to the low energy and momentum resolution of \textbf{Q}-EELS, the theoretically predicted linear dispersion and the massless exciton have yet to be observed \cite{Hong2020, Hong2021}.

In this paper, we present the first direct observation of a linear exciton dispersion, in freestanding monolayer hexagonal boron nitride (hBN) using \textbf{Q}-EELS in a scanning transmission electron microscope (STEM). The linear dispersion regime in hBN is substantially larger than in TMDs due to the larger magnitude of the long-range exchange \cite{Cudazzo2016, Koskelo2017}, which facilitates the observation of the linear bands, and subsequently the non-analytic discontinuity hosting massless excitons. Our measured exciton dispersion is in excellent agreement with finite-momentum \textit{ab initio} $GW$ plus Bethe Salpeter equation ($GW$-BSE) calculations, allowing us to trace the origin of different bands observed in experiment and identify the lowest exciton energy band from the pristine monolayer hBN. We find that the dipole-allowed direct optical band gap in monolayer hBN is 6.6 eV, considerably larger than the previous estimation from optical spectroscopy \cite{elias2019direct} and theory \cite{Cudazzo2016, Koskelo2017, Zhang2022}. Our $GW$-BSE calculation also reveals a dipole-forbidden indirect exciton band that resides slightly below or above the energy of the direct optical gap, whose exact positions depends on strain and the choice of exchange-correlation functional. These findings significantly advance our understanding of 2D excitons and open up possibilities for engineering long-lived coherent exciton states and their application in optoelectronic technologies.

\begin{figure}
 \centering
 \includegraphics[width=0.65\linewidth]{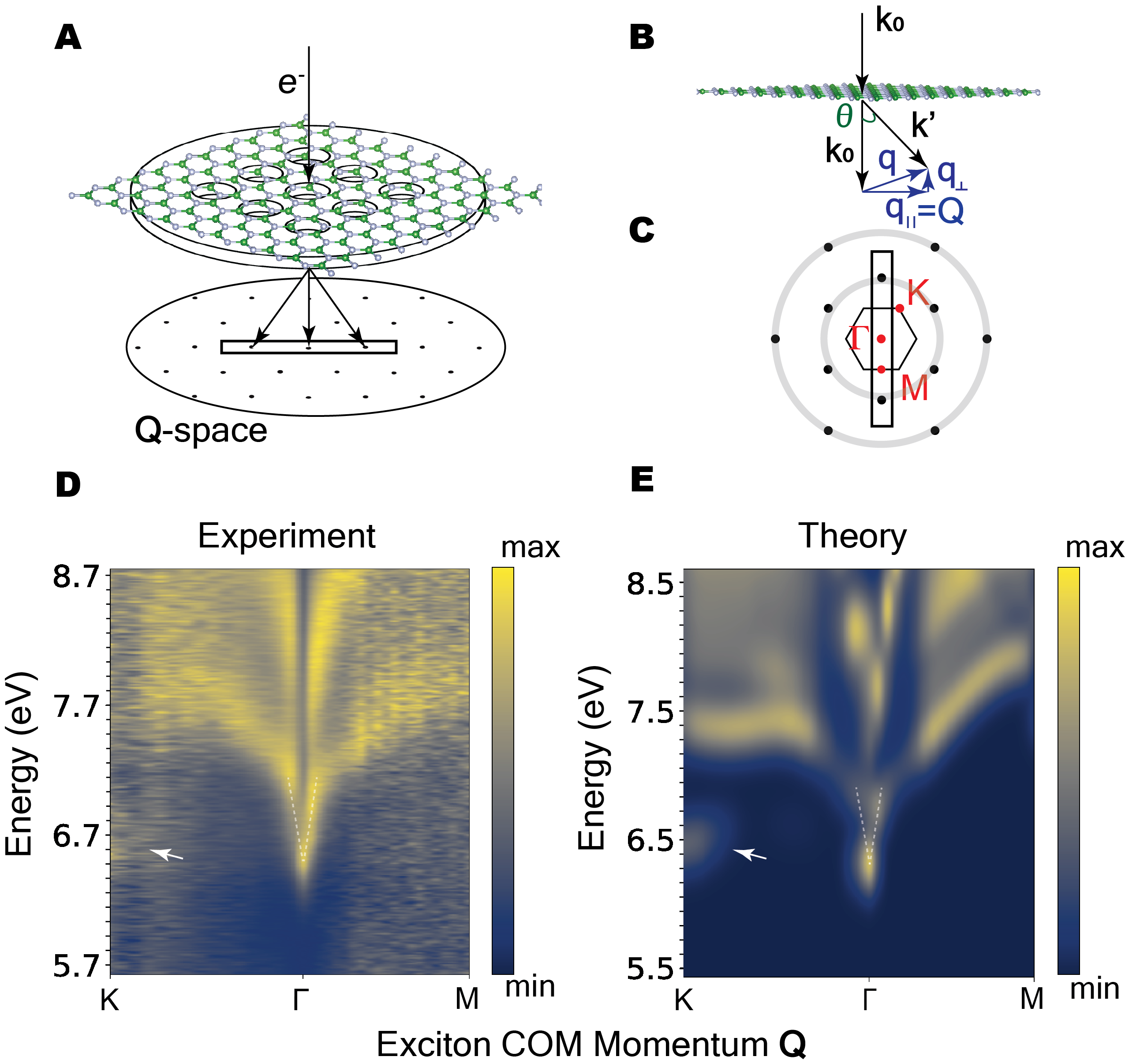}
 \caption{\textbf{Momentum-resolved electron energy-loss spectrum for freestanding monolayer hBN.} (\textbf{A}) Scattering geometry for momentum-resolved EELS (\textbf{Q}-EELS) in STEM mode (convergent beam). Electron beam ($e^-$) perpendicular to the freestanding monolayer hBN is scattered into the \textbf{Q}-space. A slot aperture in \textbf{Q}-space is used for selecting a path in \textbf{Q}-space. (\textbf{B}) Definition of the scattering vector $\textbf{q} = \text{\textbf{k}}_0 - \text{\textbf{k}}'$, where $\text{\textbf{k}}_0$ and $\text{\textbf{k}}'$ are the wave vectors of the incident and inelastically scattered electrons, respectively, and $\theta$ is the scattering angle. The components $\textbf{q}_{\parallel}$ and $\textbf{q}_{\perp}$ denote the in-plane and out-of-plane components of $\textbf{q}$, where $\textbf{q}_{\parallel}$ = \textbf{Q}. (\textbf{C}) Schematic of the electron diffraction pattern and the first Brillouin zone (BZ) of hBN in \textbf{Q}-space. An example of the placement of slot aperture is shown to be aligned along the $\Gamma$--M direction. (\textbf{D}) Experimental $E$-\textbf{Q} (energy-momentum) colormap of the \textbf{Q}-EELS for freestanding monolayer hBN. (\textbf{E}) $E$-\textbf{Q} dispersion predicted by $GW$-BSE calculation. COM: center-of-mass. White dashed lines and arrows indicate the linearly dispersive bands around $\Gamma$ and the broad exciton feature at K, respectively. Note that spectra in \textbf{D} and \textbf{E} are renormalized at each \textbf{Q} (see Supplementary Materials for details).}
 \label{experiment}\
\end{figure}

\subsection*{Experimental evidence for linearly dispersive exciton band}
The experimental setup and schematic are illustrated in Figure \ref{experiment} (A) and (B). We transferred a freestanding monolayer of hBN, grown by chemical vapor deposition (CVD), onto a TEM grid with a holey carbon support. This setup enabled measurements on the suspended regions without any influence of a substrate. These measurements were conducted at liquid nitrogen temperature, focusing on the freestanding areas depicted in Figure \ref{experiment}(A). 

To determine the exciton dispersion, we used momentum-resolved EELS (\textbf{Q}-EELS) in a STEM. In a typical EELS experiment, an electron beam passes through the sample, causing some electrons to scatter inelastically and lose energy from exciting the sample, which alters both the energy ($E$) and momentum ($\mathbf{k}_0$) of the incident electrons, and a momentum $\mathbf{Q}$ is transferred to excitations in the sample, as shown in Figure \ref{experiment}(B). The EELS spectrometer then captures the distribution of these scattered electrons, producing an energy-loss spectrum averaged over a range of momentum transfers delimited by a round spectrometer entrance aperture, obscuring momentum-specific information. To achieve momentum selectivity while ensuring that each momentum transfer \textbf{Q} is obtained under identical conditions, a rectangular spectrometer entrance slot (or slit) aperture (Figure \ref{experiment}(C)) was used to isolate a specific momentum transfer direction, encompassing a large momentum range (up to the third BZ) in a single-shot acquisition. Additional details on the measurement procedures are provided in the Materials and Methods section of the Supplementary Materials, where spatially-resolved images and electron diffraction patterns are included (Figure S1).

Figure \ref{experiment}(D) presents the experimental \textbf{Q}-EELS data for the monolayer region of hBN along the K–$\Gamma$–M high-symmetry path as a colormap.  Two dispersive exciton features are clearly visible: a lower, linearly dispersive exciton band centered at $\Gamma$, and several higher bands composed of excited exciton states. The lowest-energy exciton branch shows clear linearly dispersive behavior with a non-analytical discontinuity, making it the first experimental observation of the theoretically predicted massless exciton~\cite{qiu2015nonanalyticity}. Additionally, a broad excitonic feature is observed at $\textbf{Q}=\text{K}$, marked by white arrows.

To compare with the experimental data, we also calculated the exciton band structure of monolayer hBN within the \textit{ab initio} $GW$-BSE approach \cite{deslippe2012berkeleygw, hybertsen1986electron, rohlfing2000electron} (see Supplementary Materials for computational details), as shown in Figure \ref{experiment}(E). The theoretical results closely match the experimental data, reproducing both the linearly dispersing exciton band at $\textbf{Q}=\Gamma$ , the non-analytical discontinuity, and the broad exciton feature at $\textbf{Q}=\text{K}$. The high-energy excited-state features also align well with the measurements. The significance and characteristics of these features are discussed in the following section.

\begin{figure}
 \centering
 \includegraphics[width=0.75\linewidth]{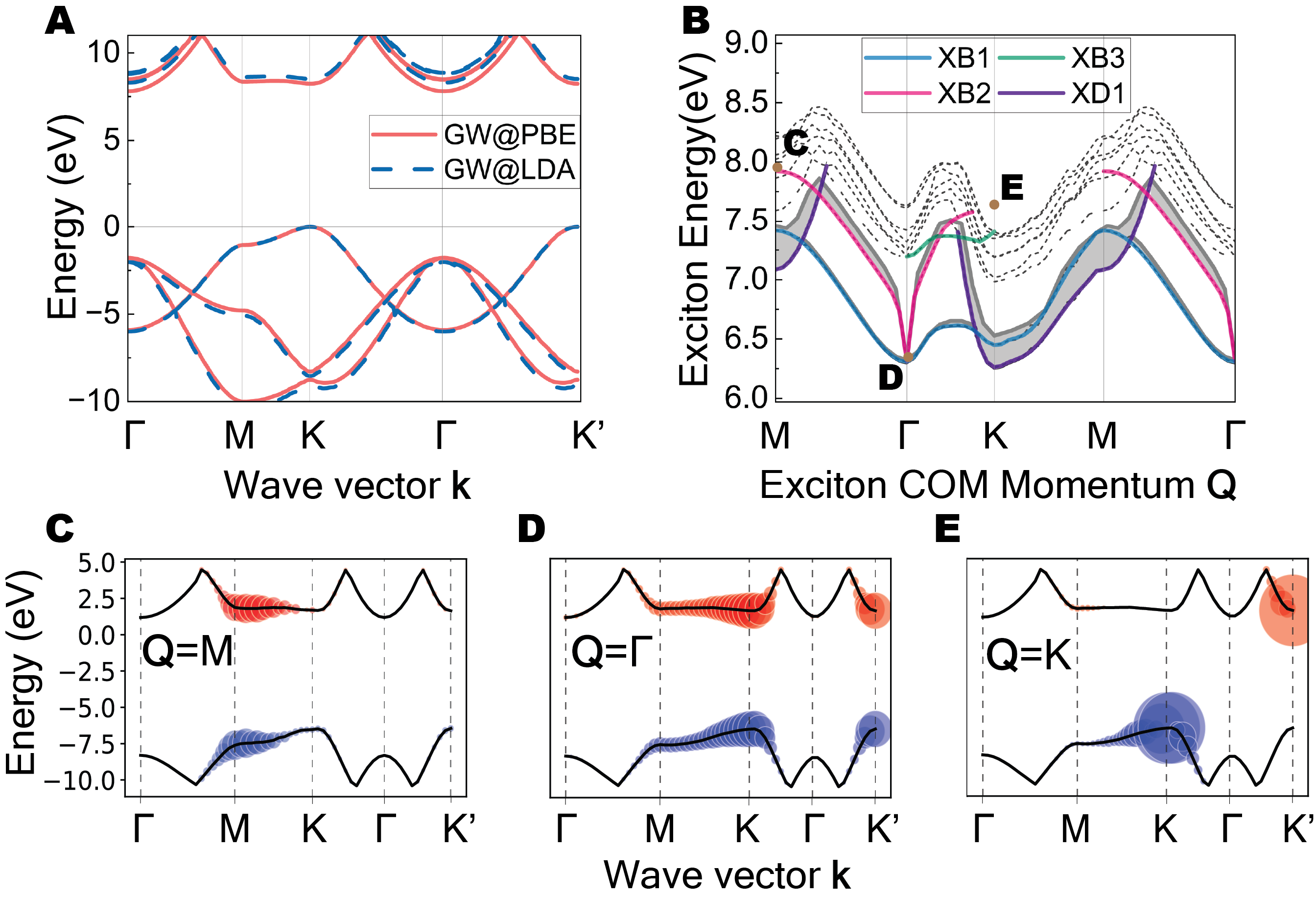}
\caption{\textbf{Electronic and excitonic band structures of monolayer hBN from first-principles calculations.} (\textbf{A}) QP band structures obtained by GW@PBE (solid coral red) and by GW@LDA (dashed blue) of monolayer hBN, with the valence band maximum set to 0 eV. (\textbf{B}) Calculated exciton band structure, displaying the dispersion of the nine lowest energy exciton states (dashed grey). Four exciton branches are highlighted---three lowest energy exciton bands (XB1, XB2, XB3) and a dipole forbidden exciton band (XD1) centered at $\textbf{Q} = K$. Points labeled C, D, E correspond to the wavefunctions shown in plots \textbf{C}--\textbf{E}. (\textbf{C}--\textbf{E}) Exciton envelope wave function projected onto the electronic band structure of hBN with momentum transfer $\textbf{Q}$ = M (\textbf{C}), $\textbf{Q} = \Gamma$ (\textbf{D}) and $\textbf{Q} =K$ (\textbf{E}) for XB2.}
 \label{bandstructure}
\end{figure}

\subsection*{Quasiparticle and exciton band structure of monolayer hBN}
Figure \ref{bandstructure} (A) shows the theoretically calculated band structure of monolayer hBN at the $GW$ quasiparticle (QP) level calculated on top of two different mean fields within density functional theory (DFT), specifically the local density approximation ($GW$@LDA)~\cite{perdew1981self} and the generalized gradient approximation of Perdew, Burke, and Ernzerhof ($GW$@PBE)~\cite{perdew1996generalized}. To remove unphysical dependence on the mean field screening, we self-consistently updated the screening in $W$ with the optical band gap (see Supplementary Materials). We find that the band gap of monolayer hBN can be either direct or indirect at the DFT level, depending on the lattice parameter and choice of exchange correlation (XC) functional (see Supplementary Materials Table S1). After incorporating dynamical many-body effects through the $GW$ self energy, the QP band gap of hBN becomes indirect, regardless of the choice of mean field, consistent with previous calculation \cite{Zhang2022}. Throughout our calculations, we used an experimental lattice parameter of a = 2.504 \r{AA} \cite{ishigami2003properties}.  At the $GW$@PBE($GW$@LDA) level, hBN has an indirect QP band gap from the valence band maximum (VBM) at K to the conduction band minimum (CBM) at $\Gamma$ of 7.91(8.29) eV and a direct band gap at K of 8.48(8.50) eV. Intriguingly, the exciton binding energy at the direct gap is significantly larger than the exciton binding energy at the indirect gap, and thus, while the QP band structure is unvaryingly indirect, the exciton band structure (Fig. \ref{bandstructure}(B)) can have a direct or indirect gap at the BSE level depending on the choice of DFT mean field. However, the lowest energy dipole-allowed exciton is always at \textbf{Q} = $\Gamma$. The lowest energy dipole-forbidden exciton, which we label XD1, has a center-of-mass (COM) momentum $\vb{Q}$ = K and involves a parity forbidden transition between nitrogen $p_z$ orbitals in the valence band at K and boron $s$ orbitals in the conduction band at $\Gamma$. XD1 has also been previously identified as a free-electron-like state \cite{Lechifflart2023, Marini_2024}. Interestingly, while XD1 is previously reported to be the lowest energy exciton state, we find that it is only the lowest energy state when the exciton band structure is calculated on a PBE mean field, where it has an excitation energy of 6.26 eV, corresponding to an exciton with a binding energy of 1.65 eV. However, when the exciton band structure is calculated on a LDA mean field, the exciton band gap becomes direct, as the excitation energy of XD1 is pushed up to 6.50 eV due to a corresponding increase in the QP gap. Thus, within our theory, there is a 0.24 eV uncertainty in the position of XD1. More details on the influence of the DFT relaxation and XC on the exciton band structure can be found in the Supplementary Materials.

The lowest energy dipole-allowed exciton, on the other hand is unchanged with the choice of XC functional. It is doubly degenerate at $\Gamma$, due to the degeneracy of the K and K$^\prime$ valleys in the electronic band structure. This direct exciton has a significantly larger exciton binding energy of 2.17(2.16) eV, on a PBE(LDA) mean field, and an optical gap of 6.31(6.34) eV, on a PBE(LDA) mean field, which lies within 50 meV (PBE) or 200 meV (LDA) of the indirect exciton, leaving open the possibility of tuning the relative energies of the direct and indirect states through substrate interactions, as suggested in previous work \cite{Lechifflart2023}, or through strain engineering (see Supplementary Materials). As the dipole-allowed exciton band moves away from $\Gamma$, the interplay of the intervalley and intravalley long-range exchange interaction \cite{qiu2015nonanalyticity} causes the exciton bands to split into a parabolically dispersing transverse branch (XB1) and a linearly dispersing longitudinal branch (XB2). The linear dispersion has equal group and phase velocities of $v_{\text{M}}=8\times 10^{5}$ m/s (0.003$c$) along the $\Gamma$--M direction and $v_{\text{K}}=6\times 10^{5}$ m/s (0.002$c$) along the $\Gamma$--K direction, which is comparable to the Fermi velocity of Dirac electrons in graphene \cite{elias2011dirac}. The higher-energy excitons at $\Gamma$, which comprise of the exciton Rydberg series, also consist of pairs of parabolically and linearly dispersing states that are degenerate at $\Gamma$. Signatures of these higher-energy states are visible in both the experimental and theoretical Q-EELS (Figure \ref{experiment}(D, E)), but their individual dispersions cannot be resolved due to the smaller energy difference between each excited state and the larger spectral weight near $\Gamma$. In addition, a band XB3 is also labeled in the calculated band structure to illustrate where band-crossings of the exciton bands occur.

To further elucidate the character of the finite-momentum excitons, we analyze the exciton envelope wavefunction $A^{S(\vb{Q})}_{vc\vb{k}}$, which describes the contribution of the valence state ($v$) and conduction state ($c$) to the exciton $|S(\vb{Q})\rangle=\sum_{vc\vb{k}} A^{S}_{vc\vb{k}}|v\vb{k},c\vb{k+Q}\rangle$. We sum the amplitudes of contributions over valence (conduction) states and project onto the conduction (valence) bands in the band structure in Figure~\ref{bandstructure} (C--E).  Focusing on the XB2 exciton, which is responsible for the observed linear dispersion, we examined the exciton wavefunctions at different momenta \textbf{Q} = M, $\Gamma$, K. The size of the colored dots on the band structure corresponds to $\sum_v \abs{A^S_{vc\vb{k}}}^2$ for the conduction states or $\sum_c \abs{A^S_{vc\vb{k}}}^2$ for the valence states. These analyses reveal that the XB2 exciton involves transitions along the high-symmetry directions corresponding to M-M$^\prime$ and K-K$^\prime$ vectors, rather than the equivalent $\Gamma$--M or $\Gamma$--K directions.

\begin{figure}
 \centering
 \includegraphics[width=0.65\linewidth]{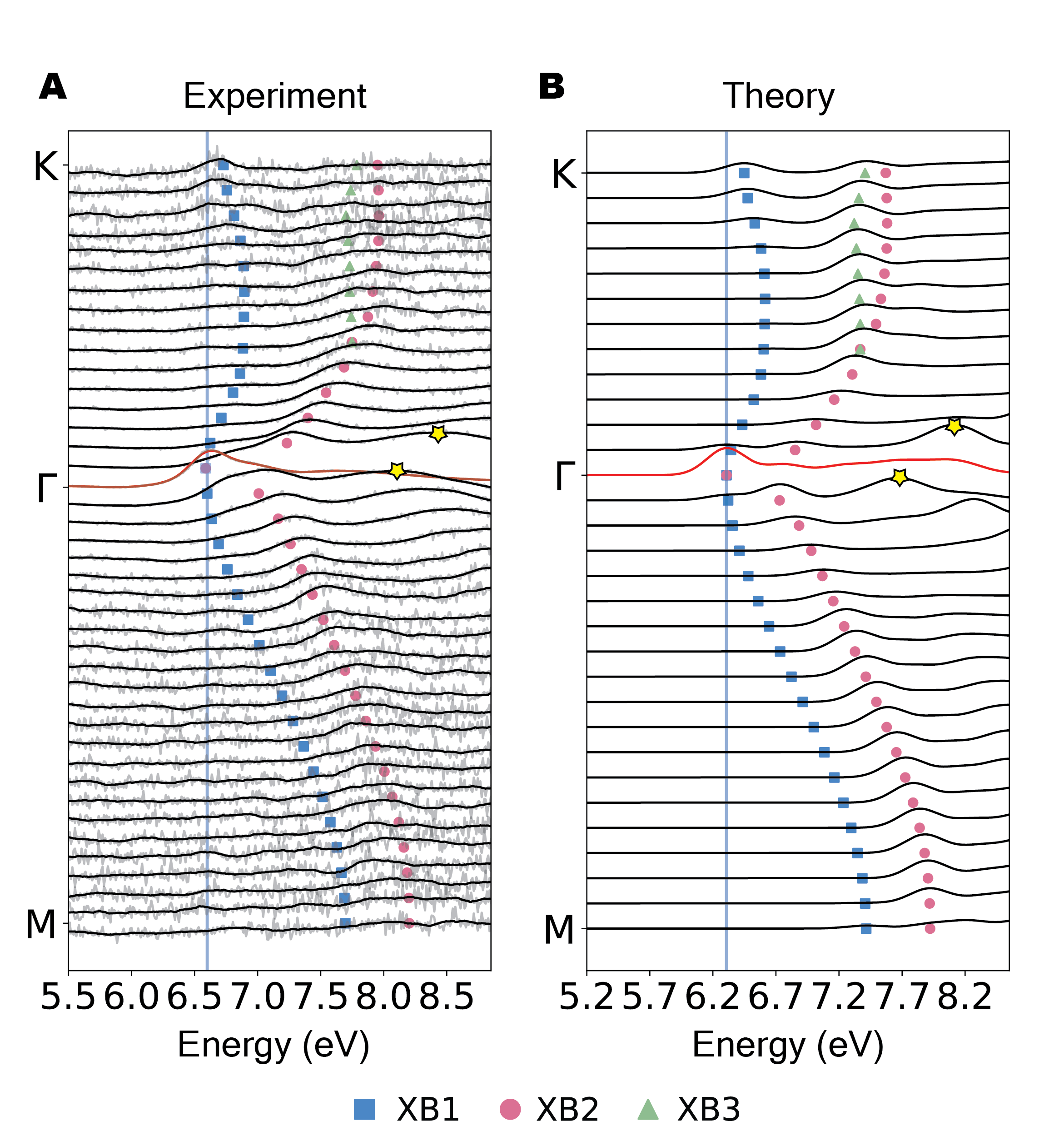}
 \caption{\textbf{Experimental and theoretical momentum-resolved electron energy-loss spectra.} (\textbf{A}) Momentum-resolved EELS data showcasing the dispersion of excitons in freestanding monolayer hBN. The gray lines are the raw data, and the black lines are the smoothed data. The three lowest energy bright exciton bands XB1 (blue), XB2 (pink), and XB3 (green) and several higher energy peaks (yellow stars) are uniformly blue-shifted by 0.3 eV and overlaid on the experimental spectrum. (\textbf{B}) Loss functions calculated from $GW$-BSE, with a uniform empirical energy broadening of 0.15 eV applied, along with the appropriate momentum broadening (see Materials and Methods). All spectra are normalized at each $\textbf{Q}$.}
 \label{EELS}
\end{figure}

\subsection*{Q-EELS}
Figure \ref{EELS}(A) shows the experimental $\textbf{Q}$-EEL spectra, marked by the bands identified in theory. The raw data are shown in light gray, while the smoothed data is plotted in black. A vertical line marks the lowest energy peak, which occurs at 6.6 eV at the $\Gamma$ point. The energy-loss function corresponds to the longitudinal component of the imaginary part of the inverse of the macroscopic dielectric function \cite{kuzmany2009solid}. To interpret the experimental results, we calculate the loss function using \textit{ab initio} finite momentum \textit{GW}-BSE (see Supplementary Materials for computational details). The corresponding theoretical spectra with a uniform empirical broadening of 0.15 eV are shown in Figure \ref{EELS}(B). As a guide to the eye, the three lowest energy exciton bands from the calculated band structure (XB1, XB2, and XB3) are uniformly blue-shifted by 0.3 eV and overlaid on the experimental results. Additional high-energy features are marked by yellow stars to facilitate comparison. XB1 corresponds to the dispersion of the lowest-energy transverse singlet exciton, and XB2 corresponds to the linearly dispersing longitudinal singlet exciton. As \textbf{Q} moves away from $\Gamma$, the abrupt change in energy of the XB2 exciton peak demonstrates its linearly dispersive nature; even with a small shift in momentum, its change in energy is very large. Although XB1 generally exhibits lower oscillator strength across most momentum transfer vectors \textbf{Q}, it shows increased oscillator strength at \textbf{Q} = K. This matches our experimental findings, which display a clear peak at \textbf{Q} = K, or the broad exciton feature mentioned in the previous section. The high-energy features, labeled with yellow stars, also agree well with each other. The excellent agreement between the experimental and theoretical spectra unambiguously reveals that the linear feature starting at 6.6 eV corresponds to the dispersion of the massless exciton on XB2. 

Intriguingly, our $\textbf{Q}$-EELS spectra indicate that the direct optical gap of hBN is 6.6 eV, consistently across four separate $\textbf{Q}$-EELS experiments on six different samples. This result contrasts with previous photoluminescence (PL) and cathodoluminescence (CL) studies, which identified the optical band gap at 6.1 eV \cite{elias2019direct,Rousseau2021, Shima_2024}. However, earlier measurements were performed on monolayer hBN supported by a substrate, such as graphite, which is known to induce significant renormalization effects \cite{ugeda2014giant, qiu2017environmental, utama2019dielectric, raja2017coulomb}, whereas our samples are freestanding. Additionally, prior studies \cite{elias2019direct, Shima_2024, Rousseau2021} did not record data above 6.2 eV due to limitations in light source and decreased detector quantum efficiency in the deep ultraviolet range. Luminescence measurements can also be influenced by defect and phonon-assisted recombination processes, whereas EELS and optical absorption/reflection measurements explicitly excite excitons. Our theoretical calculations are also in agreement with the larger experimentally observed band gap. In contrast to prior \textit{GW}-BSE calculations, which relied on screening from the underestimated DFT band gap to obtain an optical gap of 5.95 eV \cite{Zhang2022, Lechifflart2023}, we self-consistently update the screening to match the optical band gap, resulting in a theoretical optical gap of 6.23 eV in absorption and 6.31 eV in EELS (see supplemental text in Supplementary Materials).

\subsection*{Theoretical Analysis}
Next, we elucidate the origin of this linearly dispersive band (XB2). As previously predicted for TMDs~\cite{qiu2015nonanalyticity}, in 2D materials, the long-range exchange gives rise to linearly dispersing excitons, while the addition of $C_3$ symmetry results in the interplay of intervalley and intravalley long-range exchange in the K and K$^\prime$ valleys, which splits the bright degenerate exciton states into two bands: one a transverse band having parabolic dispersion (XB1), the other a longitudinal band having a V-shaped light-like linear dispersion  with a non-analaytic discontinuity at $\Gamma$ (XB2). 

However, in TMDs, this linear dispersion is limited to approximately 10\% of the Brillouin zone, making it challenging to observe in \textbf{Q}-EELS due to the limited experimental energy and momentum resolution \cite{Hong2020}. As seen in Figure \ref{bandstructure}(B), this effect is clearly much larger and more extensive in monolayer hBN. We fit the exciton dispersion to the model Hamiltonian previously proposed in Ref. \cite{qiu2015nonanalyticity}. Diagonalizing the effective Hamiltonian yields two solutions:
one with a parabolic dispersion
\begin{equation}
    \Omega_-(\textbf{Q}) = \Omega_0 + \frac{\hbar ^2 |\mathbf{\textbf{Q}}|^2}{2M_-^{*}},
\end{equation}
and the other with a non-analytic dispersion
\begin{equation}
    \Omega_+(\textbf{Q}) = \Omega_0 + 2A|\textbf{Q}| + \frac{\hbar ^2 |\mathbf{\textbf{Q}}|^2}{2M_+^{*}}.
\end{equation}
Here, $\Omega_0$ is the exciton energy at $\Gamma$, $A$ is a proportionality constant, and $M_{\pm}$ is the effective mass associated with the quadratic term in the exciton dispersion for the parabolic ($-$) branch and the linear ($+$) branch. The slope of the linear branch, denoted as $2A$, provides a measure of the strength of the long-range exchange interaction. In monolayer hBN, $A$ = 2.77 eV$\cdot$\r{AA} along the $\Gamma$--M direction and 1.94 eV$\cdot$\r{AA} along the $\Gamma$--K direction --- both values notably larger than the previously reported 0.9 eV$\cdot$\r{A} for monolayer MoS$_2$. This enhancement is due to the more ionic nature of hBN, reflected in the large electronegativity difference between B and N, compared with the atom pairings of TMDs. The more ionic nature of the B--N bond, which produces a valence band with primarily nitrogen $p_z$ character and a conduction band with primarily boron $p_z$ character, results in a larger exciton dipole moment, and hence, a stronger long-range exchange, which has the form of a dipole-dipole interaction \cite{andreani1988longitudinal, egri1985excitons}. The linear dispersion gives rise to an exciton of zero effective mass with a group velocity of 0.003$c$ along $\Gamma$--M. For the parabolic branch, the large exciton binding energy instead gives rise to a significant mass enhancement: the electron (hole) band mass is 1.16$m_e$ along K--$\Gamma$ and highly non-parabolic along the K--M direction, while the exciton mass, $M_{-}^{*}$ is renormalized to 1.93$m_e$ along $\Gamma$--M and 8.74$m_e$ along $\Gamma$--K. 

The linear exciton dispersion is notable because it gives rise to a high exciton group and phase velocity that is uniform over a large region of momentum space. In fact, the velocity of the linear exciton is considerably higher than the velocity of acoustic phonons in monolayer hBN, suggesting that the linearly dispersing exciton could have an exceptionally long coherence time due to the lack of momentum- and energy-conserving scattering channels with acoustic phonons, leading to the possibility of ultrafast, long-range quasi-ballistic exciton transport. To explore this possibility, we calculate the exciton-phonon scattering matrix elements between a finite momentum state on the linear exciton branch and the rest of the exciton band structure within first principles many-body perturbation theory \cite{antonius2022theory, chen2020exciton, chan2023exciton}. We clearly see that exciton-phonon scattering along the linearly dispersing band and between the linear and parabolic bands near $\Gamma$ is minimal, while there is considerable scattering to the rest of the Brillouin zone (Figure S9). This suggests the possibility of engineering long-lived linearly dispersing excitons by increasing the energy of XD1 through substrate or strain engineering. Further details may be found in the Supplementary Materials.

To conclude, we achieve the first direct measurement of linearly dispersing excitons in a 2D material, indicating the existence of massless excitons with the possibility of long-lived phase coherence. Our $\vb{Q}$-EELS measurements are in excellent agreement with our $GW$-BSE calculations, allowing us to unambiguously identify signatures of low-energy linearly dispersing excitons, as well as finite-momentum excitons and excitonic excited states, giving us access to the full exciton band structure and establishing $\vb{Q}$-EELS as a preeminent tool in understanding exciton dispersion, valley structures, and momentum-dark states. The linear exciton dispersion arises due to the long-range exchange interaction, which is enhanced in hBN, compared to other 2D materials, such as TMDs, as a consequence of the more ionic bonding. This discovery of a light-like exciton dispersion not only advances our fundamental understanding of 2D exciton physics but also opens new avenues for exploring exciton-mediated phenomena requiring extended phase coherence, such as superconductivity and Bose-Einstein condensation in two-dimensional systems, as well as ultrafast exciton transport in energy and optoelectronic materials. Additionally, our results illuminate longstanding questions about the nature of the band gap in monolayer hBN \cite{su2024fundamentals}, where we determine that the direct optical gap is 6.6 eV, considerably larger than previous estimates, and is in competition with a dipole-forbidden indirect gap whose relative energy can be engineered by strain or other tuning.
\clearpage
\bibliography{arxiv}
\bibliographystyle{sciencemag}

%%%%%%%%%%%%%%%% ACKNOWLEDGEMENTS %%%%%%%%%%%%%%%

\section*{Acknowledgments}
This work (LY.L., J.W., B.H., DY.Q) was primarily supported by the U.S. DOE, Office of Science, Basic Energy Sciences under Early Career Award No. DE-SC0021965. Excited-state code development was supported by Center for Computational Study of Excited-State Phenomena in Energy Materials (C2SEPEM) at the Lawrence Berkeley National Laboratory, funded by the U.S. DOE, Office of Science, Basic Energy Sciences, Materials Sciences and Engineering Division, under Contract No. DE-C02-05CH11231 (DYQ). The calculations used resources of the National Energy Research Scientific Computing (NERSC), a DOE Office of Science User Facility operated under contract no. DE-AC02-05CH11231, and and the Texas Advanced Computing Center (TACC) at The University of Texas at Austin.

Experimental \textbf{Q}-EELS research conducted as part of a user project was supported by the Center for Nanophase Materials Sciences (CNMS), which is a US Department of Energy, Office of Science User Facility at Oak Ridge National Laboratory. This research was conducted, in part, using instrumentation within ORNL's Materials Characterization Core provided by UT-Battelle, LLC under Contract No. DE-AC05-00OR22725 with the U.S. Department of Energy.

% \paragraph*{Funding:}

% \paragraph*{Author contributions:}

% \paragraph*{Competing interests:}

\paragraph*{Data and materials availability:}
The data presented in this manuscript are available from the authors upon reasonable request.

%%%%%%%%%%%%%%%% SUPPLEMENT LIST %%%%%%%%%%%%%%%
\end{document}